\documentclass[doublecol,figures]{epl2} 

\def\f#1#2{\frac{{#1}}{{#2}}}

\def\ben{\begin{eqnarray*}}
\def\een{\end{eqnarray*}}

\def\be#1{\begin{eqnarray}\label{#1}}
\def\ee{\end{eqnarray}}
\def\it#1{\textit{#1}}
\def\etc{\textit{et al. }}

\title{The gravitational bending angle by static and spherically symmetric black holes in bumblebee gravity}
\shorttitle{The gravitational bending angle by black holes in bumblebee gravity} 

\author{\'I. D. D. Carvalho\inst{1} \and G. Alencar\inst{1,2} \and W.M. Mendes\inst{3} \and R. R. Landim\inst{1}}
\shortauthor{F. Author \etal}

\institute{                    
  \inst{1} Universidade Federal do Ceará (UFC) - Departamento de Física, Campus do Pici, Fortaleza - CE, C.P. 6030, 60455-760 - Brazil\\
  \inst{2} International Institute of Physics - Federal University of Rio Grande do Norte, Campus Universitário, Lagoa Nova, Natal, RN 59078-970, Brazil\\
  \inst{3} Instituto Federal de Educação, Ciência e Tecnologia do Ceará (IFCE) - Departamento de Ensino, Campus Crateús, 63700-000, Crateús, CE, Brazil.
}

\abstract{
This work investigates the influence of the Lorentz symmetry breaking in the bending angle of massive particles and light for bumblebee black hole solutions. The solutions analyzed break the Lorentz symmetry due to a non-zero vacuum expectation value of the bumblebee field. We use the Ishihara method, which allows us to study the bending angle of light for finite distances, and it is applicable to non-asymptotically flat spacetimes when considering the receiver viewpoint. In order to analyze the deflection of massive particles, we systematize the Ishihara method for its application in the Jacobi metric. This systematization allows the study of the deflection angle of massive particles using the Gauss-Bonnet theorem. We consider two backgrounds: the first was found by Bertolami \textit{et al.} and is asymptotically flat. The second was found recently by Maluf \textit{et al.} and is not asymptotically flat due to an effective cosmological constant.}

\begin{document}

\maketitle

\section{\label{sec:level1} Introduction}

General Relativity (GR) is the classical field theory that considers gravity as a deformation of spacetime by matter and energy. This curved spacetime determines the particle trajectories \cite{weinberg1972gravitation, landau2, pad, misner1973gravitation}. This theory predicted a deflection angle of light by the Sun twice larger than the Newtonian Gravity. This led to the first of its experimental tests \cite{dyson1920determination}. As proof of its predictive power, we can mention the detection of gravitational waves in 2016 by the VIRGO and LIGO laboratories \cite{Abbott:2016blz} and the first photo of a black hole carried out by the EHT project in 2019 \cite{Akiyama:2019cqa}. The presence of a distribution of matter and energy leads to a modification of the bending of light when compared with the empty spacetime case. This bending of light by a distribution of matter and energy is called Gravitational Lens by an analogy with optics (see Refs. \cite{schneider1992gravitational, Wambsganss:1998gg, bartelmann2010gravitational} and references therein).
The study of Gravitational Lenses is very important because it is a way to investigate many subjects such as: cosmological parameters \cite{inada2012sloan, suyu2013two, Collett:2014ola}, the discovery of exoplanets \cite{Beaulieu:2006xj, gaudi2008discovery}, the investigation of massive compact halo objects (MACHOs) \cite{Alcock:2000ph}, and the use of strong gravitational lensing to investigate extra dimensions and Kalb-Ramond field \cite{Chakraborty:2016lxo}.

Beyond gravitation, the physical processes are characterized by the Standard Model (SM), which includes the electromagnetic, weak and strong interactions. This theory has a remarkable success because it is consistent with almost all experimental results \cite{Morel:2020dww} (for a historical review see\cite{Weinberg:2004kv}). However, this theory does not consider the quantum effects of gravitation, and as much as there are several candidates to explain quantum gravity, direct experiments to validate any of those theories are impractical because they are on the Plank scale ($10^{19}$ GeV). However, indirect effects can arise in low energy scales that allow the formulation of experiments. These indirect effects of quantum gravity may be associated with the violation of Lorentz Symmetry. The Standard Model Extension (SME) is an effective theory that generalizes the SM and GR, but this theory allows the violation of the Lorentz and CPT symmetries \cite{colladay1997cpt,Colladay:1998fq}. A particular SME is the bumblebee model, which violates the Lorentz symmetry and the diffeomorphism due to a vector field with a non-zero vacuum expectation value \cite{Kostelecky:2003fs, Bluhm:2004ep, bluhm2008spontaneous}. Soon later, an asymptotically flat black hole solution to the bumblebee model has been found by Bertolami and Páramos \cite{bertolami}. After this, a second static, asymptotically flat solution was found by Casana \etc \cite{casana2018exact}. These results were generalized to stationary spacetimes by Li \etc in Ref. \cite{Li:2020dln}, which represented a Kerr-like black hole solution to the bumblebee model. Finally, a static, non-asymptotically flat solution was found very recently by Maluf \etc in Ref. \cite{Maluf2020kgf}. Some studies investigate the properties of the bumblebee black holes, for example, the influence of the generalized uncertainty principle (GUP) and the Lorentz symmetry breaking (LSB) in Hawking radiation \cite{Kanzi:2019gtu}, the polarization effects in the bumblebee black hole shadow due to the coupling between the photon and the bumblebee field \cite{Chen:2020qyp}. Moreover, Ding and Chen found an exact slowly rotating Einstein-bumblebee black hole solution and investigated the LSB effects in the black hole greybody factor \cite{Ding:2020kfr}. Recently, Kanzi and Sakalli investigated the impacts of the LSB on the bosonic/fermionic quasinormal modes and greybody factors of the Kerr-like black hole in bumblebee gravity \cite{Kanzi:2021cbg}.

The Gauss-Bonnet theorem connects differential geometry with topology \cite{do2016differential, klingenberg2013course}. In 2008, Gibbons and Werner used this theorem to propose a method to compute the deflection angle of light for a static and spherically symmetric (SSS) spacetime \cite{gibbons}. In 2016, Ishihara \etc extended this approach to consider finite distances between the system receiver-lens-source \cite{ish}, this enables to investigate deflection of light for geometries that are not asymptotically flat if the source or receiver is not located at the horizon\cite{Ono:2019hkw}. Furthermore, Takizawa \etc gave another interpretation for the Ishihara method, the receiver viewpoint, where they found, in another way, the same results of Ishihara \cite{Takizawa:2020egm}. Finally, Ono \etc generalized the Ishihara method to include stationary backgrounds \cite{ono2017gravitomagnetic,ono2018deflection,ono2019deflection}. Currently, various papers use the Gauss-Bonnet theorem to investigate the deflection of light in black holes and wormholes \cite{Werner:2012rc, Jusufi:2018jof, Ovgun:2019wej, Ovgun:2018fnk, Ovgun:2018prw, Jusufi:2017drg}.

For applications of these methods to black hole backgrounds in bumblebee gravity see Refs. \cite{Ovgun:2018ran,Ovgun:2018xys}. Furthermore, in Ref. \cite{Gibbons:2015qja}, Gibbons shows that the massive particles free motion in static spacetimes is given by the geodesics of an energy-dependent Riemannian metric on the spatial sections analogous to Jacobi metric in classical dynamics. Crisnejo and Gallo were the first to study the deflection angle of massive particles in a SSS background through Jacobi's metric and the Gauss-Bonnet theorem \cite{Crisnejo:2018uyn}. Then, many works were done using the Crisnejo and Gallo approach. For example, the studying of deflection of massive particles by wormholes \cite{Li:2019vhp}, the deflection of massive particles by Kerr black hole and Teo wormholes \cite{Jusufi:2018kry}, the investigation of rotating naked singularities from Kerr-like wormholes by their deflection angles of massive particles \cite{Jusufi:2018gnz}, the studying of the circular orbit of a particle and the weak lensing effects \cite{Li:2020wvn}, and the deflection of massive particles by a Kerr-like black hole in bumblebee gravity \cite{Li:2020dln}.

In this paper, we propose to apply the Ishihara method to compute the bending of massive particles and light for the backgrounds of Refs. \cite{bertolami,Maluf2020kgf}. The structure of the paper is as follows. In the next section, we review the bumblebee model and the background solutions of Refs. \cite{casana2018exact, bertolami,Maluf2020kgf}. In the third section, we systematize the Ishihara method for applying it in the deflection of massive particles case. In the fourth section, we use the Ishihara method systematized to compute the deflection of massive particles for the Bertolami and Páramos background, and the Maluf and Neves background. Finally, in fifth section, we summarize our main results.

\section{\label{sec:level2}Black hole solutions in bumblebee gravity}

In this section, we review the bumblebee model. It is a SME that violates the Lorentz symmetry through a vector field $B_\mu$, called the bumblebee field. Although very simple, this model has interesting features as boost, rotation, and CPT violation. The nonvanishing vacuum expectation value of the bumblebee field causes the Lorentz and diffeomorphism violation \cite{Kostelecky:2003fs,Bluhm:2004ep}. The action of the bumblebee model for a spacetime with a cosmological constant and null torsion is given by
\begin{widetext}
\begin{equation}\label{2.1}
S_B =\int d^4x\sqrt{-g}\left[\f{1}{2\kappa}(R - 2\Lambda+\xi R_{\mu\nu}B^\mu B^\nu)-\f{1}{4}B_{\mu\nu}B^{\mu\nu}-V(B_\mu B^\nu \pm b^2) \right],
\end{equation}
\end{widetext}
\begin{floatequation}
\mbox{\textit{see eq.~(\ref{2.1}),}}
\end{floatequation}
where $\kappa = 8 \pi $, $\Lambda$ is a cosmological constant, $\xi$ is a real constant that controls the non-minimal coupling between bumblebee field and curvature and $b^2$ is a real positive constant. The bumblebee field strength is defined as

\be{2.2}
B_{\mu\nu}= D_\mu B_\nu - D_\nu B_\mu, 
\ee
where $D_\mu$ is the covariant derivative. The determination of the vacuum expectation value of the bumblebee field is given when the potential is zero, $V( B^{\mu} B_\mu \pm b^2) = 0$. When this occurs, we have $B^{\mu} B_\mu \pm b^2=0$, and the bumblebee field vacuum expected value is $b^\mu$, with $b^\mu b_\mu = \mp b^2$. The quantity $b_\mu$ is the coefficient of Lorentz and CTP violation, and the $\pm$ signs in the potential dictate the nature of $ b_ \mu $ as timelike or spacelike. For more details in SME and bumblebee model see Refs. \cite{colladay1997cpt,Colladay:1998fq,Kostelecky:2003fs, Bluhm:2004ep, bluhm2008spontaneous, casana2018exact, bertolami, Maluf2020kgf} and references therein. Hereafter, we will present three black hole solutions in bumblebee gravity and its characteristics. Two of these solutions \cite{bertolami, casana2018exact} are called Schwarzschild-like because they do not consider a cosmological constant ($\Lambda = 0$). These solutions become the Schwarzschild solution when the vacuum expectation value of the bumblebee field is null. The other solution is Kottler-like, which arises when we take into account a non-zero cosmological constant. This black hole solution becomes the Kottler solution when the vacuum expectation value of the bumblebee field is null. The Kottler solution describes matter from cosmological constant $\Lambda$ around a spherical source of mass. This solution generalizes the Schwarzschild vacuum solution \cite{muller2009catalogue}. The solutions mentioned above share some hypotheses. First, they are SSS solutions. Therefore, the ansatz for the metric is given by

\be{2.2.1}
g_{\mu\nu} = diag \left(- e^{2\gamma(r)},e^{2\rho(r)},r^2,r^2\sin^2{\theta}\right),
\ee
where $\gamma(r)$ and $\rho(r)$ are functions only of the radial coordinate. Second, the hypothesis that the bumblebee field $B_\mu$ remains frozen in its vacuum expectation value $b_\mu$. Moreover, it is assumed that the vacuum expectation value of the bumblebee field $b_\mu$ is radial-like, $b_\mu = (0,b_r(r),0,0)$.

\subsection{Schwarzchild-like solution}

The first solution we will discuss is the one found by Bertolami and Páramos \cite{bertolami}. In order to find $b_\mu$ they impose $D_\mu b_\nu = \partial_\mu b_\nu - \Gamma^{\sigma}_{\mu\nu}= 0$, which implies that $b_r(r)=\xi^{-1/2}b_0 e^{\rho}$. So, the  line-element for the Bertolami and Páramos solution is given by

\be{2.4}
ds^2 = - f_B(r)dt^2+\f{1}{f_B(r)}dr^2 + r^2d\Omega^2,~~
\ee
where $f_B(r) =1-\f{2M}{r}\left(\f{r}{L_0}\right)^{l_B}$. In the above equations $L_0$ is an arbitrary distance, $l_B \approx b_{0}^{2}/2$, and $M=Gm$ is the geometrical mass\cite{bertolami}. The second solution was found by Casana \etc \cite{casana2018exact}. The authors do not use the condition $D_\mu b_\nu = 0 $. The way used by them to determine $b_\mu$ is by the condition $b^\mu b_\mu = b^2 =const.$, and with this $b_r(r)= \left| b \right| e^{\rho(r)}$. So, the  line-element for the Casana \etc solution is given by

\be{2.6}
ds^2=-\left(1-\f{2M}{r}\right)dt^2+\f{(1+l_c)}{1-\f{2M}{r}}dr^2+r^2d\Omega^2,
\ee
where $M=G m$ is the geometric mass and $l_c = \xi b^2$. Therefore $l_c$ depends on the vacuum expectation value of the bumblebee field and on the coupling constant $\xi$.

\subsection{Kottler-like solution}

Now we will review a black hole solution found recently by Maluf and Neves\cite{ Maluf2020kgf}. They consider a cosmological constant in the bumblebee gravity, therefore it is not asymptotically flat. The action used to construct this solution is given in equation (\ref{2.1}). However, differently from the two solutions above, the cosmological constant is non-zero. Maluf and Neves used a linear potential

\be{2.7}
V(B_\mu B^\nu \pm b^2)=\f{\sigma}{2}\left(B_\mu B^\nu - b^2\right),
\ee 
where $\sigma$ is a Lagrange-multiplier. Note that $V = 0$ and $V'(y)\equiv dV(y)/dy = \sigma/2$ when $B_\mu B^\nu = b^2$. The linear potential was chosen to produce a black hole solution with a non-zero cosmological constant. With this they found the solution

\be{2.8}
ds^2=- f_M(r)dt^2+(1+l_c)\f{1}{f_M(r)}dr^2+r^2d\Omega^2,~~
\ee
where

\be{2.8.1}
f_M(r)= 1-\f{2M}{r}- (1+l_c)\f{\Lambda_{e}}{3}r^2.
\ee
In the above equations we have used $l_c = \xi b^2$ and $\Lambda_{e} = \kappa \sigma/\xi$. Therefore, we have an effective cosmological constant which depends on the non-minimal coupling constant $\xi$ and on the Lagrange-multiplier field $\sigma$. Moreover, the Maluf and Neves solution becomes the Casana \etc solution when we consider a null cosmological constant ($\Lambda_e = 0$).

\section{\label{sec:level5}Bending angle of massive particles in a static and spherically symmetric background}

In this section, we systematize the Ishihara method to covers the deflection of massive particles in a SSS spacetime. In Ref. \cite{Gibbons:2015qja}, Gibbons shows that the free motion of massive particles in static spacetimes is given by the geodesics of an energy-dependent Riemannian metric on the spatial sections. This assembles the Jacobi metric in classical dynamics. Crisnejo and Gallo were the first ones to study the deflection angle of massive particles in a {SSS} background through Jacobi's metric and the Gauss-Bonnet theorem \cite{Crisnejo:2018uyn}. However, as far as we know, this generalization has not been systematized, and we do it here. First, the Jacobi metric of a SSS spacetime is given by 

\begin{equation}\label{5.0}
ds^2 = (E^2 - m^2 A(r))\left[\frac{B(r)}{A(r)}dr^2+\frac{r^2}{A(r)}d\phi^2\right],
\end{equation}
where we are considering the equatorial plane $\theta = \pi/2$ \cite{Gibbons:2015qja}. By using equation (\ref{5.0}), we get the angular momentum conserved due to axial symmetry, given by
\begin{equation}\label{5.1}
J \equiv (E^2 - m^2 A(r))\frac{r^2}{A(r)}\frac{d\phi}{ds} = constant.
\end{equation}
The trajectory equation for a massive particle is then obtained by using equations (\ref{5.0}) and (\ref{5.1}), and we find 
\begin{equation}\label{5.2}
(E^2 - m^2 A(r))^2\frac{B(r)}{A(r)} \left(\frac{dr}{ds}\right)^2 = E^2 - \left(m^2 + \frac{J^2}{r^2}\right)A(r).
\end{equation}
We define the impact parameter as $\mathrm{b}$. We consider the coordinate transformation $r=1/u$, $J = mv\mathrm{b}/\sqrt{1-v^2}$ and $E = m/\sqrt{1-v^2}$ \cite{Crisnejo:2018uyn}. Then, equation (\ref{5.2}) becomes
\begin{equation}\label{5.3}
\left(\frac{du}{d\phi}\right)^2 = \frac{1}{A B}\left[\frac{1}{\mathrm{b}^2v^2}-\left(\frac{1-v^2}{\mathrm{b}^2v^2}+ u^2\right)A\right].
\end{equation}
We can rewrite equation (\ref{5.0}) using $E^2 = m^2/(1-v^2)$. By doing that, we get
\begin{equation}\label{5.4}
ds^2 = m^2 \left(\frac{1}{1-v^2} - A(r)\right)\left[\frac{B(r)}{A(r)}dr^2+\frac{r^2}{A(r)}d\phi^2\right].
\end{equation}
For more details in the SSS Jacobi metric see Ref. \cite{Gibbons:2015qja}. We are ready to study the gravitational bending angle of massive particles in the SSS background, considering finite distance. We define 
\begin{equation}\label{5.5}
ds^2 = \Omega^2\left[\frac{B}{A}dr^2+ \frac{r^2}{A}d\phi^2\right],
\end{equation}
where $\Omega^2 \equiv m^2(1/(1-v^2) - A)$. First, we consider the unit vectors at the radial direction and angular direction of the SSS Jacobi space. The unit radial vector is $e_{rad}= (\sqrt{A}/(\sqrt{B}\Omega),0)$ and the unit angular vector is $e_{ang}=(0, \sqrt{A}/(\Omega r))$. The unit vector tangent to the trajectory is $(K^r,K^\phi)= (dr/ds,d\phi/ds)$. We can rewrite $K$ by using the equations \begin{equation}\label{5.8}
(K^r,K^\phi) = \frac{J}{\Omega^2}\frac{A}{r^2}\left(\frac{dr}{d\phi},1\right).
\end{equation}

\begin{figure}
  \onefigure[width=5cm,height=4cm]{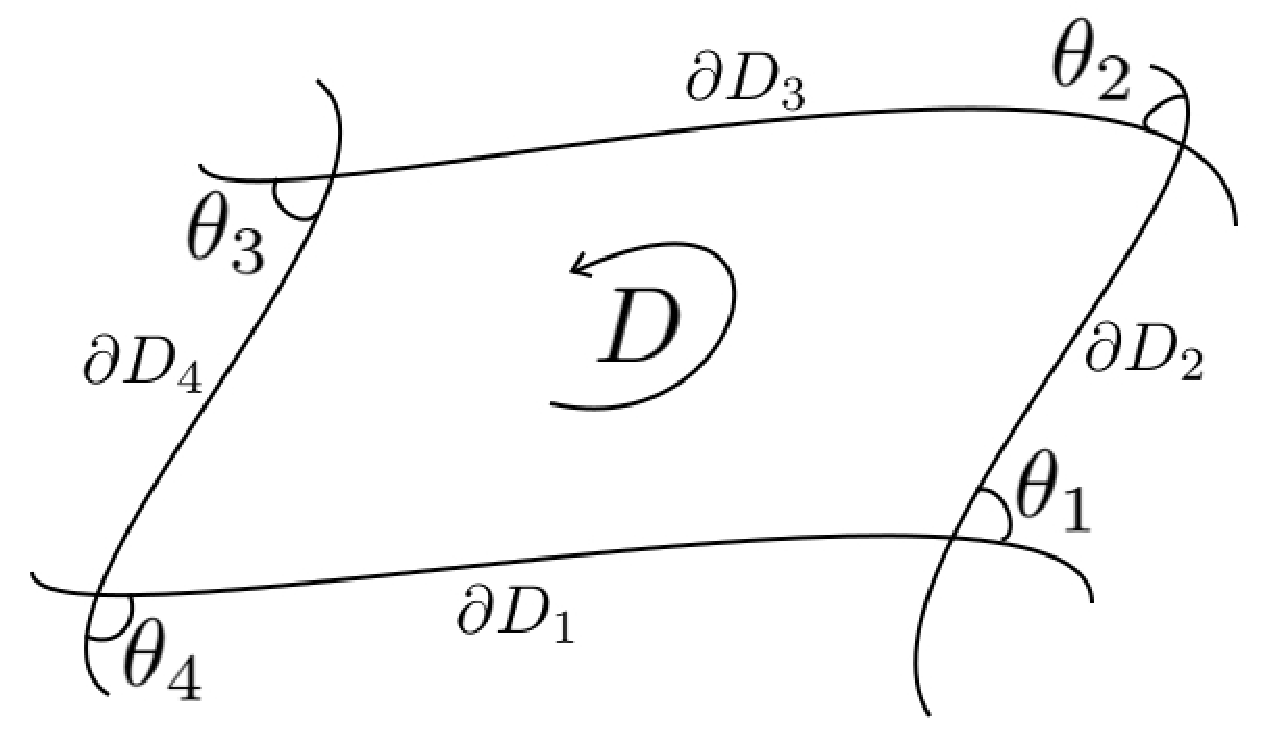}
  \caption{Schematic figure for the Gauss-Bonnet theorem.}
  \label{fig1}
\end{figure}

The Gauss-Bonnet theorem associates the geometrical characteristics of a surface with its topological Euler characteristic. Let the domain ($D$,$\chi$,$g$) be a subset of a compact, oriented surface, with Euler characteristic $\chi$ and a Riemann metric $g$ giving rise to a Gaussian curvature $K$. Moreover, let $\partial D: \{t\}\rightarrow D$ be a piecewise smooth boundary with geodesic curvature $\kappa$, and $\theta_i$ the external angle at $i$th vertex, traversed in the positive sense(see figure \ref{fig1}). Then the Gauss-Bonnet theorem can be stated as

\be{3.0}
\int\int_{D} K dS + \int_{\partial D} \kappa dt + \sum_i \theta_i = 2 \pi \chi(D). 
\ee
For more details about the Gauss-Bonnet theorem and its applications see Refs.\cite{do2016differential,klingenberg2013course}.
The Ishihara method uses the Gauss-Bonnet theorem to find the deflection angle of light, this method provides the deflection angle of light as 

\be{3.5}
\alpha = \Psi_R - \Psi_S + \int_{u_R}^{u_0}\frac{du}{\sqrt{F(u)}}+\int_{u_S}^{u_0}\frac{du}{\sqrt{F(u)}}.
\ee
To compute the integrations present in equation (\ref{3.5}), we will use the second fundamental theorem of calculus, that is
\be{4.5}
\int_{u}^{u_0} \f{du'}{\sqrt{F(u')}}= I(u_0) - I(u),
\ee
where $I(u)$ is the integral of $F(u)^{-1/2}$. For the case of the deflection angle of massive particles, let us build $\Psi$ and $F(u)$ in the SSS Jacobi space. First, we consider $\Psi$ defined as the angle between the unit tangent vector and the unit radial vector at the trajectory point. Now, by applying the inner product between the unit tangent vector and the unit radial vector, we find 
\begin{equation}\label{5.10}
\cos{\Psi} = \Omega^2 \left(\frac{B}{A}K^r e_{rad}^r + \frac{r^2}{A}K^\phi e_{rad}^\phi\right) = \sqrt{AB}\frac{J}{\Omega r^2}\frac{dr}{d\phi}.
\end{equation}
If we use the fundamental relation of trigonometry,we avoid the rate $dr/d\phi$. Then
\begin{equation}\label{5.11}
\sin^2{\Psi} = 1 - \frac{AB}{\Omega^2}J^2\left(\frac{du}{d\phi}\right)^2,
\end{equation}
where we have used $r=1/u$. By using equation (\ref{5.3}), we obtain
\begin{equation}\label{5.12}
\Psi = \arcsin{\left(\mathrm{b}uv \sqrt{\frac{A}{1- A (1-v^2)}}~\right)}.
\end{equation}
In the case of the deflection of light, we use $v=1$. This means that in equation (\ref{5.12}) the value of $\Psi$ is the same that the one found in the Ishihara method \cite{ish}. Finally, we can find the $F(u)$ present in equation (\ref{3.5}). We use equation (\ref{5.3}) because of the definition $(du/d\phi)^2 \equiv F(u)$. Then for massive particles in a SSS background, $F(u)$ reads 
\begin{equation}\label{5.13}
F(u) = \frac{1}{A B}\left[\frac{1}{\mathrm{b}^2v^2}-\left(\frac{1-v^2}{\mathrm{b}^2v^2}+ u^2\right)A\right].
\end{equation}
We should point that the case of deflection of light can be obtained in the limit $v\to 1$. With this, equation (\ref{5.13}) becomes the definition of $F(u)$ present in the Ishihara method \cite{ish}. Then the deflection angle computed by using equations (\ref{5.12}) and (\ref{5.13}) at (\ref{3.5}) agrees with the Ishihara method and covers the deflection angle for massive and massless particles in SSS spacetime.

\section{\label{sec:level6}Bending angle of massive particles for black hole solution in bumblebee gravity}

This section is designated to compute the deflection angle of massive particles and light. To do it we use the systematization of the Ishihara method presented in the previous section. This is done for the following background: Bertolami-Paramos and Maluf-Neves. We investigate the influence of the bumblebee field considering finite distances of the Receiver-Lens-Source system.

\subsection{Bertolami's Schwarzschild-like background}\label{Subsec-3-2}

The Bertolami and Páramos background is expressed by equation (\ref{2.4}). We can study the deflection angle of massive particles in the Bertolami and Páramos background through equation (\ref{3.5}). Making the following identifications: For $\Psi$ we use equation (\ref{5.12}), and for $F(u)$ we use equation (\ref{5.13}). First, we can find $\Psi_R - \Psi_S$ by an expansion of equation (\ref{5.12}) until first order of $M/\mathrm{b}$, where we get
\begin{eqnarray}\label{EQ-3-1-2}
\Psi_R - \Psi_S &=& \arcsin{\mathrm{b}u_R} + \arcsin{\mathrm{b}u_S} - \pi \nonumber \\
&-&\frac{M}{\mathrm{b}v^2}\left(\frac{\mathrm{b}}{L_0}\right)^{l_B} \frac{\mathrm{b}^2u_S^2}{\sqrt{1-\mathrm{b}^2u_S^2}} + O\left(\frac{M^2}{\mathrm{b}^2}\right). \nonumber \\
\end{eqnarray}\label{EQ-3-1-3}
Second, we can compute $I(u)$ for the Bertolami and Páramos background:
\begin{eqnarray}\label{EQ_3-1-3}
I(u) &=& \arcsin{\mathrm{b}u} - \frac{M}{\mathrm{b}v^2}\left(\frac{\mathrm{b}}{L_0}\right)^{l_B}\frac{1+v^2 - v^2\mathrm{b}^2u^2}{\sqrt{1-\mathrm{b}^2u^2}}\nonumber \\
&+& O\left(\frac{M^2}{\mathrm{b}^2}\right).
\end{eqnarray}
Then the deflection angle of massive particles in Bertolami and Páramos background reads 
\begin{eqnarray}\label{EQ-3-1-4}
\alpha &=& \frac{M}{\mathrm{b}}\left(\frac{1+v^2}{v^2}\right)\left(\frac{\mathrm{b}}{L_0}\right)^{l_B}\left[\sqrt{1-\mathrm{b}^2u_S^2}+\sqrt{1-\mathrm{b}^2u_R^2}\right] \nonumber \\
&+& O\left(\frac{M^2}{\mathrm{b}^2}\right).
\end{eqnarray}
Now, we will analyze the deflection angle of massive particles, equation (\ref{EQ-3-1-4}), in three limiting cases. 

First, note that in the case of the deflection angle of light ($v=1$), equation (\ref{EQ-3-1-4}) becomes 
\begin{eqnarray}\label{4.8}
\alpha \approx \f{2M}{\mathrm{b}}\left(\f{\mathrm{b}}{L_0}\right)^{l_B}\left[\sqrt{1-\mathrm{b}^2u_R^2}+\sqrt{1-\mathrm{b}^2u_S^2}\right].
\end{eqnarray}
Second, note that disregarding finite distances ($\mathrm{b}u_R\approx \mathrm{b}u_S \approx 0$), equation (\ref{EQ-3-1-4}) becomes
\begin{equation}\label{EQ-3-1-5}
\alpha \approx \frac{2M}{\mathrm{b}}\left(\frac{1+v^2}{v^2}\right)\left(\frac{\mathrm{b}}{L_0}\right)^{l_B}.
\end{equation}
Finally, considering a null vacuum expectation value of the bumblebee field ($l_B= 0$), equation (\ref{EQ-3-1-5}) becomes 
\begin{equation}\label{5.19}
\alpha  \approx \frac{2M}{\mathrm{b}}\left(\frac{1+v^2}{v^2}\right).
\end{equation}
The result of equation (\ref{5.19}) agrees with Ref. \cite{Crisnejo:2018uyn}. This is expected because when ($l_B= 0$) the Bertolami and Páramos background becomes the Schwarzschild background.

\subsection{Maluf Kottler-like background}\label{Subsec-3-3}
The Maluf and Neves background is expressed by equations (\ref{2.8}) and (\ref{2.8.1}). Let us compute the deflection angle of massive particles in such background. First, we can find $\Psi_R - \Psi_S$ by an expansion of equation (\ref{5.12}) until first order of $M/\mathrm{b}$ and $(1+l_c) \Lambda_e$, obtaining

\begin{eqnarray}\label{EQ-4-31}
  \Psi_R - \Psi_S &=& \arcsin{\mathrm{b}u_R} + \arcsin{\mathrm{b}u_S} - \pi \nonumber \\
  &-&\frac{M}{\mathrm{b}v^2}\left[\frac{\mathrm{b}^2u_R^2}{\sqrt{1-\mathrm{b}^2u_R^2}}+\frac{\mathrm{b}^2u_R^2}{\sqrt{1-\mathrm{b}^2u_R^2}}\right] \nonumber \\
  &-& \frac{\lambda^2 \Lambda_e \mathrm{b}^2}{6v^2}\left[\frac{(\mathrm{b}u_R)^{-1}}{\sqrt{1-\mathrm{b}^2u_R^2}}+\frac{(\mathrm{b}u_S)^{-1}}{ \sqrt{1-\mathrm{b}^2u_S^2}}\right] ~~\nonumber \\
  &+& \frac{\lambda^2 \Lambda_e M\mathrm{b}}{6v^2} \left[\frac{3-4v^2+2(2v^2-1)\mathrm{b}^2u_R^2}{(1-\mathrm{b}^2u_R^2)^{3/2}}\right. \nonumber \\
  &+&\left.\frac{3-4v^2+2(2v^2-1)\mathrm{b}^2u_R^2}{(1-\mathrm{b}^2u_R^2)^{3/2}}\right] \nonumber \\
  &+& O\left(\ \f{M^2}{\mathrm{b}^2},\lambda^2\Lambda_e^2\right),
\end{eqnarray}
where $\lambda \equiv \sqrt{1 + l_c}$. Second, we can compute $I(u)$ for the Maluf background, finding

\begin{eqnarray}\label{EQ-4-32}
  I(u) &=& \lambda \arcsin{\mathrm{b}u} - \frac{M}{\mathrm{b}v^2}\frac{\lambda (1+v^2)-\lambda v^2\mathrm{b}^2u^2}{\sqrt{1-\mathrm{b}^2u^2}} \nonumber \\
  &-&\frac{\lambda^2 \Lambda_e \mathrm{b}^2}{6v^2}~\frac{\lambda(v^2-1)+\lambda(2-v^2)\mathrm{b}^2u^2}{\mathrm{b}u \sqrt{1-\mathrm{b}^2u^2}} \nonumber \\
  &-& \frac{\lambda^2 \Lambda_e M\mathrm{b}}{6v^2} ~\frac{2(2v^2 - 1)\lambda -3\lambda v^2\mathrm{b}^2u^2}{(1-\mathrm{b}^2u^2)^{3/2}}\nonumber \\
  &-& \frac{\lambda^3 \Lambda_e M\mathrm{b}}{6} \left(\frac{1-v^2}{v^2}\right)^2 \frac{3\mathrm{b}^2u^2-4}{(1-\mathrm{b}^2u^2)^{3/2}}\nonumber \\
  &-& \frac{\lambda^3 \Lambda_e M\mathrm{b}}{2} \left(\frac{1-v^2}{v^2}\right)^2 \tanh^{-1}{(\sqrt{1-\mathrm{b}^2u^2})} \nonumber \\
  &+& O\left( \f{M^2}{\mathrm{b}^2},\lambda^2\Lambda_e^2\right),
\end{eqnarray}
where $\tanh^{-1}(x)$ is the inverse hyperbolic tangent. The deflection angle of massive particles in this background is
\begin{widetext}
\begin{eqnarray}\label{EQ-4-33}
  \alpha &=& (1-\lambda) \left[\arcsin{ \mathrm{b} u_R} + \arcsin{\mathrm{b}u_S} - \pi\right] \nonumber \\
  &+&\frac{M}{\mathrm{b}v^2}\left[\frac{\lambda (1+v^2)-(1+\lambda v^2)\mathrm{b}^2u_R^2}{\sqrt{1-\mathrm{b}^2u_R^2}}+\frac{\lambda (1+v^2)-(1+\lambda v^2)\mathrm{b}^2u_R^2}{\sqrt{1-\mathrm{b}^2u_R^2}}\right] \nonumber \\
  &+& \frac{\lambda^2 \Lambda_e \mathrm{b}^2}{6v^2} \left[~\frac{\lambda(v^2-1 + (2-v^2)\mathrm{b}^2u_R^2)-1}{\mathrm{b}u_R \sqrt{1-\mathrm{b}^2u_R^2}}+\frac{\lambda(v^2-1 + (2-v^2)\mathrm{b}^2u_R^2)-1}{\mathrm{b}u_S \sqrt{1-\mathrm{b}^2u_R^2}}~\right] \nonumber \\
  &+&\frac{\lambda^2 \Lambda_e M\mathrm{b}}{6v^2} \left[\frac{4v^2(\lambda-1)+3-2\lambda+(4v^2-3\lambda v^2-2)\mathrm{b}^2u_R^2}{(1-\mathrm{b}^2u_R^2)^{3/2}}+\frac{4v^2(\lambda-1)+3-2\lambda+(4v^2-3\lambda v^2-2)\mathrm{b}^2u_R^2}{(1-\mathrm{b}^2u_R^2)^{3/2}}\right]\nonumber \\
  &+&\frac{\lambda^3 \Lambda_e M\mathrm{b}}{6} \left(\frac{1-v^2}{v^2}\right)^2 \left[\frac{3\mathrm{b}^2u_R^2-4+3(1-\mathrm{b}^2u_R^2)^{3/2}\tanh^{-1}{(\sqrt{1-\mathrm{b}^2u_R^2})}}{(1-\mathrm{b}^2u_R^2)^{3/2}} \right. \nonumber \\
  &+& \left. \frac{3\mathrm{b}^2u_R^2-4+3(1-\mathrm{b}^2u_R^2)^{3/2}\tanh^{-1}{(\sqrt{1-\mathrm{b}^2u_R^2})}}{(1-\mathrm{b}^2u_R^2)^{3/2}}\right] + O\left(\ \f{M^2}{\mathrm{b}^2},\lambda^2\Lambda_e^2\right).
\end{eqnarray}
\end{widetext}
\begin{floatequation}
\mbox{\textit{see eq.~(\ref{EQ-4-33}).}}
\end{floatequation}

Now, we will analyze the deflection angle of massive particles, equation (\ref{EQ-4-33}), in three limit cases. 

First, note that in the case of the deflection angle of light ($v=1$), equation (\ref{EQ-4-33}) becomes
\begin{eqnarray}\label{4.23}
\alpha &\approx& (\lambda-1)\left[\pi -\arcsin{(\mathrm{b}u_R)}-\arcsin{(\mathrm{b}u_R)}\right] \nonumber\\
&+&\f{M}{\mathrm{b}}\left[ \f{2\lambda - \mathrm{b}^2u_S^2(\lambda+1)}{\sqrt{1-\mathrm{b}^2u_S^2}}+ \f{2\lambda - \mathrm{b}^2u_R^2(\lambda+1)}{\sqrt{1-\mathrm{b}^2u_R^2}}\right] \nonumber\\
&-& \f{\lambda^2\Lambda_e \mathrm{b}^2}{6}\left[\f{1-\lambda \mathrm{b}^2u_R^2}{\mathrm{b}u_R\sqrt{1-\mathrm{b}^2u_R^2}} + \f{1-\lambda \mathrm{b}^2u_S^2}{\mathrm{b}u_S \sqrt{1-\mathrm{b}^2u_S^2}}\right]\nonumber \\
&+& \f{\lambda^2\Lambda_e M\mathrm{b}}{6}\left[~\f{2\lambda -1 +(2-3\lambda)\mathrm{b}^2u_R^2}{\left(1-\mathrm{b}^2u_R^2\right)^{3/2}}+\right. \nonumber \\
&+&\left. \f{2\lambda -1 +(2-3\lambda)\mathrm{b}^2u_S^2}{\left(1-\mathrm{b}^2u_S^2\right)^{3/2}}~\right].
\end{eqnarray}
Second, note that disregarding finite distances ($\mathrm{b}u_R\approx \mathrm{b}u_S \approx 0$), equation (\ref{EQ-4-33}) becomes
\begin{eqnarray}\label{EQ-4-34}
\alpha &\approx& (\lambda-1)\pi +\frac{\lambda 2M}{\mathrm{b}}\left(\frac{1+v^2}{v^2}\right)\nonumber \\
&+& \frac{\lambda^2 \Lambda_e \mathrm{b}^2}{6v^2} \left[~\frac{\lambda(v^2-1)-1}{\mathrm{b}u_R}+\frac{\lambda(v^2-1)-1}{\mathrm{b}u_S}~\right] \nonumber \\
&+&\frac{\lambda^2 \Lambda_e M\mathrm{b}}{3v^2} \left[4v^2(\lambda-1)+3-2\lambda\right]\nonumber \\
&+&\frac{\lambda^3 \Lambda_e M\mathrm{b}}{12} \left(\frac{1-v^2}{v^2}\right)^2 \left[3\pi-16 \right].
\end{eqnarray}
Finally, considering a null vacuum expectation value of the bumblebee field ($\lambda = 1$), equation (\ref{EQ-4-34}) becomes 

\begin{eqnarray}\label{EQ-4-35}
\alpha &\approx& \frac{2M}{\mathrm{b}}\left(\frac{1+v^2}{v^2}\right)+\frac{ \Lambda_e \mathrm{b}^2}{6v^2} \left[~\frac{v^2-2}{\mathrm{b}u_R}+\frac{v^2-2}{\mathrm{b}u_S}~\right] \nonumber \\
&+&\frac{\lambda^2 \Lambda_e M\mathrm{b}}{3v^2} +\frac{ \Lambda_e M\mathrm{b}}{12} \left(\frac{1-v^2}{v^2}\right)^2 \left[3\pi-16 \right].
\end{eqnarray}

The apparent divergences presented in equations (\ref{EQ-4-34}) and (\ref{EQ-4-35}) is problematic only in the case that the source or receiver is located at the horizon \cite{Ono:2019hkw}. Moreover, we must point that we can find the deflection angle of massive particles for the Casana background taking a null cosmological constant. Using $\Lambda_e = 0$ in equation (\ref{EQ-4-33}), the deflection angle of massive particles for Casana background is 
\begin{eqnarray}\label{5.17}
\alpha &=& (1-\lambda) \left[\arcsin{\mathrm{b}u_R} + \arcsin{\mathrm{b}u_S} - \pi\right] \nonumber \\
&+&\frac{M}{\mathrm{b}v^2}\left[\frac{\lambda (1+v^2)-(1+\lambda v^2)\mathrm{b}^2u_S^2}{\sqrt{1-\mathrm{b}^2u_S^2}} \right.\nonumber \\
&+&\left. \frac{\lambda (1+v^2)-(1+\lambda v^2)\mathrm{b}^2u_R^2}{\sqrt{1-\mathrm{b}^2u_R^2}}\right]+O\left(\frac{M^2}{\mathrm{b}^2}\right).
\end{eqnarray}
The equation (\ref{5.17}) agrees with Ref. \cite{Li:2020dln} and with Ref. \cite{Li:2020wvn}.

\section{\label{sec:conclusion}Conclusion}

In this paper, we computed the deflection angle of massive particles by black hole backgrounds in bumblebee gravity. We use the Ishihara method. Which can be used in non asymptotically flat spacetimes when we consider the receiver viewpoint. In the third section, we systematize the Ishihara method, and its application for massive particles. Furthermore, we analyze the influence of the bumblebee field on this deflection. We considered two backgrounds. The first one was found by Bertolami and Páramos in Ref. \cite{bertolami}. It is asymptotically flat, and becomes the Schwarzschild solution when the bumblebee field has a null vacuum expectation value. The other one was found by Maluf and Neves in Ref. \cite{Maluf2020kgf}. It is not asymptotically flat due to an effective cosmological constant. For each background, we also investigated the deflection of massive particles in three limiting cases: a) the light deflection case ($v = 1$), b) when both the receiver and the source are very far from the lens ($\mathrm{b}u_R \approx \mathrm{b}u_S \approx 0$), c) when considering the null vacuum expectation value of the bumblebee field ($\lambda = 1$ or $l_B=0$).

We found the deflection angle of massive particles in the Bertolami and Páramos background. Considering finite distances in equation (\ref{EQ-3-1-4}). We analyzed the cases a), b) and c) for this background. In the case a), when we considered the light deflection ($v=1$), equation (\ref{EQ-3-1-4}) became equation (\ref{4.8}). In case b), when both the receiver and the source are very far from the lens ($\mathrm{b}u_R \approx \mathrm{b}u_S \approx 0$), equation (\ref{EQ-3-1-4}) became equation (\ref{EQ-3-1-5}). In the case c), when we considered the null vacuum expectation value of the bumblebee field ($l_B = 0$), equation (\ref{EQ-3-1-5}) became (\ref{5.19}). This happens because if the Bertolami background have a null vacuum expectation value of the bumblebee field, the background becomes Schwarschild like.

We also analyzed the influence of the Lorentz symmetry break on the deflection angle of massive particles in the Maluf and Neves background. Furthermore, the cosmological constant depends on the characteristics of the bumblebee field interaction ($\Lambda_{e} = \kappa \sigma/\xi$). The deflection angle of massive particles considering finite distances, for this background, was presented in equation (\ref{EQ-4-33}). We analyzed the cases a), b) and c) in such configuration. In the case a), when considering the light deflection ($v=1$), equation (\ref{EQ-4-33}) became equation (\ref{4.23}). In case b), when both the receiver and the source are very far from the lens ($\mathrm{b}u_R \approx \mathrm{b}u_S \approx 0$), equation (\ref{EQ-4-33}) became equation (\ref{EQ-4-34}). In the case c), when we consider the null vacuum expectation value of the bumblebee field ($\lambda = 1$), quation (\ref{EQ-4-34}) became equation (\ref{EQ-4-35}). As in the case of light deflection, the apparent divergences presented in equations (\ref{EQ-4-34}) and (\ref{EQ-4-35}) is problematic only in the case that the source or receiver is located at the horizon, see Refs. \cite{Ono:2019hkw,Takizawa:2020egm}. We found the deflection angle of massive particles for the Casana background. We did it considering a null cosmological constant, that we showed in equation (\ref{5.17}). In the near future, we intend to study the deflection of massive particles by using the Ishihara method in the wormhole backgrounds.

\acknowledgments
We thanks Coordenação de Aperfeiçoamento de Pessoal de Nível Superior (CAPES), and Conselho Nacional de Desenvolvimento Científico e Tecnológico (CNPq) for the financial support.

\end{document}